\documentclass[showkeys,showpacs,amssymb,amsmath,preprint]{revtex4}
\usepackage{amsmath}
\usepackage{amssymb}
\usepackage{graphicx}
\usepackage{dcolumn}
\usepackage{bm}
\usepackage[colorlinks=false,dvipdfm]{hyperref} 
\usepackage{setspace}

\newcommand{\tr}{\mbox{Tr}}
\begin{document}

\author{Zhihao Ma}
\affiliation{Department of Mathematics, Shanghai Jiaotong
University, Shanghai, 200240, P.R.China\\PHONE:
011+8621-5474-4334-1774, FAX: 011+8621-5474-3152}

\author{Fu-Lin Zhang}
\affiliation{Theoretical
Physics Division, Chern Institute of Mathematics, Nankai University,
Tianjin, 300071, P.R.China
\\PHONE: 011+8622-2350-9287,
FAX: 011+8622-2350-1532}

\author{Jing-Ling Chen}
\email[Email:]{chenjl@nankai.edu.cn}\affiliation{Theoretical Physics
Division, Chern Institute of Mathematics, Nankai University,
Tianjin, 300071, P.R.China
\\PHONE: 011+8622-2350-9287,
FAX: 011+8622-2350-1532}

\date{\today}

\begin{abstract}
Fidelity plays an important role in quantum information theory. In
this letter, we introduce new metric of quantum states induced by
fidelity, and connect it with the well-known trace metric, Sine
metric and Bures metric for the qubit case. The metric character is
also presented for the qudit (i.e., $d$-dimensional system) case.
The CPT contractive property and joint convex property of the metric
are also studied.
\end{abstract}

\title{Fidelity induced distance measures for quantum states}
\pacs{03.67.-a, 03.65.Ta} \keywords{ Fidelity; Metric; Quantum
state} \maketitle

\section{Introduction}

Suppose one has two quantum states $\rho$ and $\sigma$, then the
fidelity \cite{Fid1,Fid2} between $\rho$ and $\sigma$ is given by
\begin{eqnarray} \label{ConP}
F(\rho, \sigma)=[\tr
\sqrt{\rho^{\frac{1}{2}}\sigma\rho^{\frac{1}{2}}}]^{2}
\end{eqnarray}

Fidelity plays an important role in quantum information theory and
quantum computation \cite{Niebook}, and it has deep connection with
quantum entanglement \cite{Ma064305}, quantum chaos \cite{Zana0903},
and quantum phase transitions \cite{Wang1,Wang2,Wang3}. However,
fidelity by itself is not a metric. It is a measure of the
``closeness'' of two states.
As a metric defined on quantum states, $d(x,y)$ is a function
satisfies the following four axioms:

(M1). $d(x,y) \ge 0$ for all states $x$ and $y $;

(M2). $d(x,y)=0$ if and only if  $x=y$;

(M3). $d(x,y)=d(y,x)$ for all states $x$ and $y$;

(M4). The triangle inequality: $d(x,y)\leq d(x,z)+d(y,z)$ for all
states $x, y$ and $z$.

One may expect that a metric, which is a measure of distance, can be
built up from  fidelity. Indeed, the following three functions
\begin{eqnarray}A(\rho,\sigma)&:=&\arccos{\sqrt{F(\rho,\sigma)}},\nonumber\\
B(\rho,\sigma)&:=&\sqrt{2-2\sqrt{F(\rho,\sigma)}},\nonumber\\
C(\rho,\sigma)&:=&\sqrt{1-F(\rho,\sigma)}\nonumber,
\end{eqnarray}
exhibit such metric properties. They are now commonly known in the
literature as the {\em Bures angle}, the {\em Bures metric}, and the
\emph{Sine metric} \cite{Nie062310,Ra06,Ra02}, respectively. Based
on fidelity, one can generally define a metric $D(\rho,\sigma)$ as
$D(\rho,\sigma):=\phi(F(\rho,\sigma))$, where $\phi(t)$ is a
monotonically decreasing function of $t$, and $\phi(F(\rho,\sigma))$
is required to satisfy the axioms M1-M4. From this way, one can
define many useful metrics \cite{Nie062310,Ra06,Ra02}. All of the
above three metrics belong to this type and play important roles in
quantum information theory.

The purpose of this letter is to introduce a new way to define
metric of quantum state based on fidelity. In Sec. II, the new
metric is defined. We study the qubit case in detail and naturally
connect the new metric with the well-known trace metric, the Sine
metric and Bures metric. In Sec. III, we show that the new metric
defined is truly a metric, i.e., it satisfies the axioms M1-M4. The
upper bound for the metric is also presented. Conclusion and
discussion are made in the last section.

\section{Metric induced by fidelity}

Let us define a metric of quantum states as follows:
\begin{eqnarray}D_{T}(\rho,
\sigma)=\max\limits_{\tau}|F(\rho,\tau)-F(\sigma,\tau)|\label{T-metric}
\end{eqnarray}
where the maximization is taken over all quantum states $\tau$
(mixed or pure). We call this metric $D_{T}(\rho, \sigma)$ as the
T-metric, and the state $\tau$ that attained the maximal is called
the optimal state for the metric $D_{T}(\rho, \sigma)$. The T-metric
was first introduced in \cite{Be76} for pure sates of an abstract
transition probability space (T means transition probability), but
in this letter, we define it for arbitrary quantum states in the
Hilbert space.

The above definition of metric may be not easy to calculate. If
$\tau$ is a pure state, then fidelity can be simplified as
$F(\rho,\tau)=\tr(\rho \tau)$, hence one can define another version
of metric as follows: \begin{eqnarray}D_{PT}(\rho,
\sigma)=\max\limits_{\tau}|F(\rho,\tau)-F(\sigma,\tau)|,
\end{eqnarray}
where the maximization is taken over all pure states $\tau$. We call
this metric $D_{PT}(\rho, \sigma)$ as the PT-metric, and call the
pure state $\tau$ that attained the maximal as the optimal pure
state.

In this section we consider the case of qubits (two-dimensional
quantum system). From the Bloch sphere representation, a qubit is
described by a density matrix as
\begin{eqnarray}\rho(\textbf{u})=\frac{1}{2}(\textbf{I}+\sigma\cdot
\textbf{u})\nonumber
\end{eqnarray}
 where $\textbf{I}$ is the $2\times 2$ unit matrix and
$\sigma=(\sigma_{1},\sigma_{2},\sigma_{3})$ are the Pauli matrices.
Assume $\rho(\textbf{u})$ and $\rho(\textbf{v})$ are two two states
of a qubit, then they can represented by two vectors $\textbf{u}$
and $\textbf{v}$ in the Bloch sphere. The Euclidean metric between
vectors $\textbf{\textbf{u}}$ and $\textbf{\textbf{v}}$ is defined
by
$|\textbf{u}-\textbf{v}|=\sqrt{(u_{1}-v_{1})^{2}+(u_{2}-v_{2})^{2}+(u_{3}-v_{3})^{2}}$.
The trace metric between $\rho(\textbf{u})$ and $\rho(\textbf{v})$
satisfies $D_{tr}(\rho(\textbf{u}),
\rho(\textbf{v}))=\frac{1}{2}|\textbf{u}-\textbf{v}|$, which is
proportional to the Euclidean metric.

For the qubit case, it is well-known that the fidelity has an
elegant form(\cite{Hubner92,Fid2}):
\begin{eqnarray}
F(\rho(\textbf{u}), \rho(\textbf{v}))=\frac{1}{2}[1+\textbf{u}\cdot
\textbf{v}+\sqrt{1-|\textbf{u}|^{2}}\sqrt{1-|\textbf{v}|^{2}}]\nonumber
\end{eqnarray}
where $\textbf{u}\cdot \textbf{v}$ is the inner product of two
vectors $\textbf{u}$ and $\textbf{v}$, and $|\textbf{u}|$ is the
magnitude of $\textbf{u}$. Then we have the following two Theorems.


\emph{Theorem 1.} For the qubit case,
$D_{PT}(\rho(\textbf{u}),\rho(\textbf{v}))$ equals to the trace
metric, namely
$D_{PT}(\rho(\textbf{u}),\rho(\textbf{v}))=\frac{1}{2}|\textbf{u}-\textbf{v}|
=D_{tr}(\rho(\textbf{u}),\rho(\textbf{v}))$.

\emph{Proof.} Let $\rho=\rho(\textbf{u})$, $\sigma=\rho(\textbf{v})$
and $\tau=\rho(\textbf{w})$, since $\tau$ is a pure state, which
means $|\textbf{w}|=1$, then we get
\begin{eqnarray}
|F(\rho,\tau)-F(\sigma,\tau)|&=
&\frac{1}{2}|(\textbf{u}-\textbf{v})\cdot \textbf{w}|\nonumber\\
&\leq &\frac{1}{2}|\textbf{u}-\textbf{v}|\nonumber
\end{eqnarray}
The optimal pure state is $\tau=\rho(\textbf{w})$, where
$\textbf{w}$ is a vector that parallels to $\textbf{u}-\textbf{v}$.
Thus
$D_{PT}(\rho(\textbf{u}),\rho(\textbf{v}))=\frac{1}{2}|\textbf{u}-\textbf{v}|
=D_{tr}(\rho(\textbf{u}),\rho(\textbf{v}))$.

\emph{Theorem 2.} For the qubit case,
$D_{T}(\rho(\textbf{u}),\rho(\textbf{v}))$ equals to the Sine
metric, namely $D_{T}(\rho, \sigma)=\sqrt{1-F(\rho,
\sigma)}=C(\rho,\sigma)$.

\emph{Proof.} Let $\rho=\rho(\textbf{u})$, $\sigma=\rho(\textbf{v})$
and $\tau=\rho(\textbf{w})$, then one obtains
\begin{eqnarray}
&&|F(\rho,\tau)-F(\sigma,\tau)| \nonumber\\
&&=\frac{1}{2}\biggr|(\textbf{u}-\textbf{v})\cdot
\textbf{w}+\sqrt{1-|\textbf{w}|^{2}}(\sqrt{1-|\textbf{u}|^{2}}-\sqrt{1-|\textbf{v}|^{2}})\biggr|\nonumber\\
&&\leq\frac{1}{2}\biggr[|\textbf{u}-\textbf{v}||\textbf{w}|+
\sqrt{1-|\textbf{w}|^{2}}\;|\sqrt{1-|\textbf{u}|^{2}}-\sqrt{1-|\textbf{v}|^{2}}|\biggr]\nonumber\\
&&\leq\frac{1}{2}\sqrt{|\textbf{u}-\textbf{v}|^2+
|\sqrt{1-|\textbf{u}|^{2}}-\sqrt{1-|\textbf{v}|^{2}}|^2}\nonumber\\
&&=\sqrt{1-F(\rho(\textbf{u}),\rho(\textbf{v}))}.
\end{eqnarray}

The optimal state is $\tau=\rho(\textbf{w})$, where $\textbf{w}$ is
a vector that parallels to $\textbf{u}-\textbf{v}$, and
$|\textbf{w}|=\frac{|\textbf{u}-\textbf{v}|}{\sqrt{1-F(\rho(\textbf{u}),\rho(\textbf{v}))}}$.
Thus we have $D_{T}(\rho, \sigma)=\sqrt{1-F(\rho,
\sigma)}=C(\rho,\sigma)$.

This significant result seems surprising, since we know that
$\sqrt{1-F(\rho,\sigma)}$ is the Sine metric introduced in
\cite{Nie062310,Ra06,Ra02}, which plays an important role in quantum
information processing, but here we can recover it for the qubit
case through the definition (\ref{T-metric}). One may wonder whether
the Bures metric can be obtained by the similar definition. The
answer is positive. By using the same approach developed in
\emph{Theorem 2}, for the qubit case one can prove that Bures metric
$B(\rho,\sigma)=\sqrt{2-2\sqrt{F(\rho,\sigma)}}$ can be expressed in
the following equivalent form
\begin{eqnarray}
B(\rho, \sigma)&=&\max\limits_{\tau}\;
[|\sqrt{1+|F(\rho,\tau)-F(\sigma,\tau)|}\nonumber\\
&&-\sqrt{1-|F(\rho,\tau)-F(\sigma,\tau)|}|],
\end{eqnarray}
where the maximization is taken over all pure and mixed quantum
states $\tau$.

\section{Metric character of $D_{PT}$ and $D_{T}$}

We come to discuss the case of qudit(i.e., $d\times d$ quantum
states). In this case, if $\tau$ is a pure state, then the fidelity
may have a simple form: $F(\rho,\tau)=\tr(\rho \tau)$, so we first
show the metric character of $D_{PT}(\rho, \sigma)$, where the
optimal state $\tau$ is restricted to pure state, and then turn to
show the metric character of $D_{T}(\rho, \sigma)$.

We need the following concepts: For two quantum state $\rho$ and $
\sigma$, let $\lambda_{i}$, $(i=1, 2, 3, ... , d)$, be all
eigenvalues of $\rho-\sigma$, and $\lambda_{i}$'s are arranged as
$\lambda_{1}\geq \lambda_{2}\geq...\geq \lambda_{d}$. Similarly, let
$\lambda^{'}_{i}$ be all eigenvalues of $\sigma-\rho$. Define
$E(\rho, \sigma):=\max \lambda_{i}$ and define $E(\sigma,
\rho):=\max \lambda^{'}_{i}$, so we know that $\lambda_{1}=\max
\lambda_{i}$.

Now we give an interpretation of $E(\rho, \sigma)$.  Let $\rho$ and
$\sigma$ be two quantum states, then the following is well known(for
example, see \cite{Horn91}):
\begin{eqnarray}
E(\rho, \sigma)=\max\limits_{\tau}\tr[\tau(\rho-\sigma)],
\end{eqnarray}
where the maximization is taken  over all pure states $\tau$.

Note that in general $E(\rho, \sigma)$ is not a metric, since
$E(\rho, \sigma)$ may not equal to $E(\sigma, \rho)$, but we can
symmetrize it as:
\begin{eqnarray}
D_{S}(\rho, \sigma):=\max[E(\rho, \sigma),E(\sigma, \rho)]=\max
|\lambda_{i}| ,
\end{eqnarray}
where $|\lambda_{i}|$ is the absolute value of $\lambda_{i}$. From
the knowledge of matrix analysis, $D_{S}(\rho, \sigma)$ equals to
the spectral metric between $\rho$ and $\sigma$, which was defined
as the largest singular value of $\rho-\sigma$, hence we know that
$D_{S}(\rho, \sigma)$ is nothing but the spectral metric. For the
qubit case, the $D_{S}(\rho, \sigma)$ or the spectral metric is also
equal to the trace metric, i.e.,
$D_{S}(\rho(\textbf{u}),\rho(\textbf{v}))=\frac{1}{2}|\textbf{u}-\textbf{v}|$.
Now we begin to show the metric character of $D_{PT}(\rho, \sigma)$.

\emph{Proposition 1.} For quantum states $\rho$ and $\sigma$, we
have $D_{PT}(\rho,\sigma)=D_{S}(\rho,\sigma)$, i.e, the PT-metric is
in fact the same as the spectral metric.

\emph{Proof.} From the definition $D_{PT}(\rho,
\sigma)=\max\limits_{\tau}|F(\rho,\tau)-F(\sigma,\tau)|$, since
$\tau$ is an arbitrary pure state, we have $F(\rho,\tau)=\tr(\rho
\tau)$. It yields $|F(\rho,\tau)-F(\sigma,\tau)|=|\tr(\rho
\tau)-\tr(\sigma \tau)|$, then we get $\max\limits_{\tau}|\tr(\rho
\tau)-\tr(\sigma \tau)|=\max(E(\rho, \sigma),E(\sigma,
\rho))=D_{S}(\rho,\sigma)$. The Proposition is proved.

Now we know that the PT-metric equals to the spectral metric, so it
is a true metric. In the following we shall prove that the T-metric
is also a true metric.

\emph{Theorem 3.} The T-metric $D_{T}(\rho, \sigma)$ as shown in Eq.
(\ref{T-metric}) is truly a metric, i.e, it satisfies conditions
M1-M4.

\emph{Proof.} From the definition, it is easy to prove conditions M1
and M3 hold. What we need to do is to prove conditions M2 and M4. If
$\rho=\sigma$, then of course $D_{T}(\rho, \sigma)=0$. If
$D_{T}(\rho, \sigma)=0$, we will prove $\rho=\sigma$. From the
definition, we know that $D_{T}(\rho, \sigma)\geq D_{PT}(\rho,
\sigma)$, so we get $D_{PT}(\rho, \sigma)=0$, since $D_{PT}(\rho,
\sigma)$ is a true metric, we get $\rho=\sigma$. Now we come to
prove M4, the triangle inequality $D_{T}(\rho, \sigma)\leq
D_{T}(\rho, \tau)+ D_{T}(\sigma, \tau)$. $D_{T}(\rho,
\sigma)=\max\limits_{\tau}|F(\rho, \tau)-F(\sigma, \tau)|$, and
suppose $\tau$ is the state that attains the maximal, so
$D_{T}(\rho, \sigma)=|F(\rho, \tau)-F(\sigma, \tau)|$. We assume
that $|F(\rho, \tau)-F(\sigma, \tau)|= F(\rho, \tau)-F(\sigma,
\tau)$, then we get $F(\rho, \tau)-F(\sigma, \tau)= F(\rho,
\tau)-F(w,\tau)+F(w,\tau)-F(\sigma, \tau)\leq |F(\rho,
\tau)-F(w,\tau)|+|F(w,\tau)-F(\sigma, \tau)|\leq D_T(\rho,w)+
D_T(w,\sigma)$. Thus one finally has $D_T(\rho, \sigma) \leq
D_T(\rho,w)+ D_T(\sigma, w)$.

For the qudit case, one does not have the relation $D_T(\rho,
\sigma) =\sqrt{1-F(\rho,\sigma)}$ as in \emph{Theorem 2}. However,
the numerical computation indicates the following upper bound holds:
\begin{eqnarray}
D_T(\rho, \sigma) \leq \sqrt{1-F(\rho,\sigma)}.
\end{eqnarray}
Now we will give the rigorous proof: as was shown in \cite{Ra06},
following inequality holds:
\begin{eqnarray}|F(\rho, \tau)-F(\sigma, \tau)|\leq
\sqrt{1-F(\rho,\sigma)}\end{eqnarray} for arbitrary quantum states
$\rho, \sigma,\tau$. Taking maximal in the left hand of inequality
(9), we get the inequality (8).

\section{Discussion}

In summary, we have introduced metric of quantum states induced by
fidelity, and connected it with the well-known trace metric, Sine
metric and Bures metric for the qubit case. The metric character of
$D_{T}(\rho, \sigma)$ is also presented for the qudit case.

Let us make one more discussion to end the paper.

In quantum information theory, a quantum operation or a quantum
channel is representated by a completely positive trace preserving
(CPT) map. We say that the metric $d(\rho, \sigma)$ has {\bf the CPT
 contractive property}, if $d(\phi(\rho),
\phi(\sigma))\leq d(\rho, \sigma)$ always holds, where $\phi$ is a
quantum operation, $\rho, \sigma$ be two arbitrary density matrices.
We wish that a metric is contractive under CPT map (i.e., quantum
operation), this has a physical interpretation \cite{Nie062310}: a
quantum process acting on two quantum states cannot increase their
distinguishability.

Also, we wish a metric satisfying the following {\bf joint convex
property}:  for all $0\leq \lambda \leq 1$, states
$\rho_{1},\rho_{2},\sigma_{1},\sigma_{2}$, $d(\lambda
\rho_{1}+(1-\lambda)\rho_{2}, \lambda
\sigma_{1}+(1-\lambda)\sigma_{2})\leq \lambda
d(\rho_{1},\sigma_{1})+(1-\lambda) d(\rho_{2},\sigma_{2})$.

The joint convex property also has a physical interpretation
\cite{Nie062310}: the distinguishability between the states $\lambda
\rho_{1}+(1-\lambda)\rho_{2}$ and $\lambda
\sigma_{1}+(1-\lambda)\sigma_{2}$, where $\lambda$ is not known, can
never be greater than the average distinguishability when $\lambda$
is known.

The numerical method (\cite{Bru}) show that the T-metric
$$D_{T}(\rho,\sigma)=\max\limits_{\tau}|F(\rho,\tau)-F(\sigma,\tau)|$$ {\bf satisfying} the CPT contractive property.

How about the joint convex property? We find that, the T-metric
$D_T$ is {\bf not} joint convex. However, numerical experiment shows
that its square is joint convex, that is, the following holds:
$$D^2_{T}(\lambda
\rho_{1}+(1-\lambda)\rho_{2}, \lambda
\sigma_{1}+(1-\lambda)\sigma_{2})\leq \lambda
D^2_{T}(\rho_{1},\sigma_{1})+(1-\lambda)
D^2_{T}(\rho_{2},\sigma_{2})$$

All the above evidences show that metric $D_{T}(\rho,\sigma)$ is a
useful metric in quantum information theory.

\vskip 0.05 in

{\bf Acknowledgments.}

\vskip 0.05 in

The authors wish to express their thanks to the referee for the
valuable comments and suggestions. Z.H.Ma is supported by the New
teacher Foundation of Ministry of Education of P.R.China (Grant No.
20070248087), partially supported by a grant of science and
technology commission of Shanghai Municipality (STCSM, No.
09XD1402500). J.L.C is supported in part by NSF of China (Grant No.
10605013), and Program for New Century Excellent Talents in
University, and the Project-sponsored by SRF for ROCS, SEM.

\end{document}